\documentclass[aps, twocolumn, showpacs]{revtex4}
\usepackage{pslatex}
\usepackage{epsfig}
\usepackage{dcolumn}

\newcommand{\per}[1]{#1\%}

\begin{document}
\preprint{LBNL-52051}
\title{High precision measurement of the static dipole polarizability of cesium}

\author{Jason M. Amini}
 \email{JAMaddi@lbl.gov}
 \altaffiliation[Also at ]{Physics Department, University California at Berkeley, Berkeley, CA 94720.}
\author{Harvey Gould}
 \email{Gould@lbl.gov}
\affiliation{Mail stop 71-259, Lawrence Berkeley National Laboratory,  Berkeley, CA 94720}


\date{\today}

\begin{abstract}
The cesium $6^2S_{1/2}$ scalar dipole polarizability $\alpha_0$ has been determined from the time-of-flight of laser cooled and launched cesium atoms traveling through an electric field. We find $\alpha_0 = 6.611\pm{0.009}\times 10^{-39}~\mbox{Cm}^2/\mbox{V}= 59.42\pm{0.08}\times10^{-24}~\mbox{cm}^3=401.0\pm0.6~a_0^3$.  The \per{0.14} uncertainty is a factor of fourteen improvement over the previous measurement.  Values for the $6^2P_{1/2}$ and $6^2P_{3/2}$ lifetimes and the $6^2S_{1/2}$ cesium-cesium dispersion coefficient $C_6$ are determined from $\alpha_0$ using the procedure of Derevianko and Porsev [Phys. Rev. A {\bf 65}, 053403 (2002)].
\end{abstract}


\maketitle

\begin{figure}
\epsfig{file=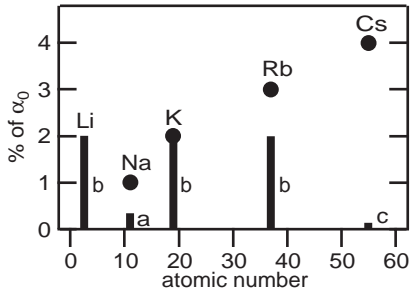}
\caption{\label{contrib} 
Comparison of the core electron contribution to the polarizability with the experimental uncertainty in the measurement of the polarizability. The fraction of the polarizability  for Na, K, Rb, and Cs  arising from the core electrons \cite{Derevianko99PRL82p3589} are shown as dots and the experimental uncertainties in the polarizability measurements are shown as bars (francium has not been measured yet).   Measurements (a) and (b)  are Refs.~\cite{Ekstrom95PRA51p3883} and \cite{Molof74PRA10p1131},  respectively, and (c) is this work.  The contribution to the polarizability from the core electrons is expected to be smaller in Li than in Na.}
\end{figure}

The static polarizability quantifies the effect of one of  the simplest perturbations to an atom:  the application of a static electric field inducing a dipole moment  \cite{Miller77AAMP13p1, Bonin97Example}.  With increasing atomic number, relativistic effects \cite{Kello93PRA47p1715,Lim99PRA60p2822} and core electron contributions \cite{Derevianko99PRL82p3589,Zhou89PRA40p5048} to the alkali polarizabilities become increasingly significant. In cesium, the heaviest stable alkali,  the relativistic effects reduce the polarizability by \per{16} and the core contributes \per{4}.    However, experimental uncertainties have made the measurements of the alkali polarizabilities relatively  insensitive to the smaller core contribution,  as shown in Fig~\ref{contrib}.  With the largest relativistic correction and core contribution of the stable alkali atoms, the cesium polarizability is an ideal benchmark for testing the theoretical treatment of both relativistic effects and core contributions.  Our measurement advances the accuracy of the cesium polarizability by a factor of fourteen over the previous measurement \cite{Molof74PRA10p1131} and places the uncertainty at \per{4} of the core contribution.

%
%


To our measurement's  level of accuracy, the hyperfine levels of the cesium ground state have a common polarizability $\alpha_0$.  From angular momentum relations \cite{Angel68PRSA305p125}, the dependencies of the polarizability on the hyperfine level $F$ and on the magnetic sublevel $M_F$ are greatly suppressed in the cesium ground state. The small remaining dependencies, generated by the hyperfine interaction, have been measured in Refs.~\cite{Gould69PR188p24,Ospelkaus03PRA67p011402R,Simon98PRA57p436}.

For a static electric field $E$ of moderate strength, the potential energy $W$ of a neutral cesium atom in that field  may be written in terms of $\alpha_0$  as $W = -(1/2) \alpha_0 E^2$. All odd terms in $E$ are disallowed by parity conservation and the linear term, also forbidden by time-reversal invariance, is experimentally known to be  less than $1.6 \times 10^{-44}$~C-m in cesium \cite{Murthy89PRL63p965}. The hyperpolarizability  contribution, which scales as $E^4$, has been calculated  \cite{Fuentealba93JPB26p2245, Palchikov00} and is  negligible at the fields used for our measurement.


Prior to the interferometric measurement of Ekstrom et. al. \cite{Ekstrom95PRA51p3883},  the most accurate determinations of the alkali  polarizabilities \cite{Hall74PRA10p1141, Molof74PRA10p1131}  had been made by measuring the deflection of a thermal beam  due to a transverse electric field gradient. The gradient generates a force ${\bf F} = - {\bf\nabla} W = \alpha_0 E {\bf\nabla} E$ where $E$ is the magnitude of the electric field.  The high velocity of a thermal beam results in a short interaction time and, consequently, a small deflection.

For cesium, we have measured instead the effect of an electric field gradient on the longitudinal velocity of a slow beam ($<2$~m/s) of neutral cesium atoms afforded by an optically launched atomic fountain.    Upon entering an electric field from a region of zero field, the kinetic energy of the atoms gain energy by the amount $-W$, producing a noticeable increase in the atoms' velocity for even moderate electric fields.  Because the force is conservative, the final velocity is dependent on only the final magnitude of the electric field and not on the details of the gradient in the region of transition from zero field.

\begin{figure}
\epsfig{file=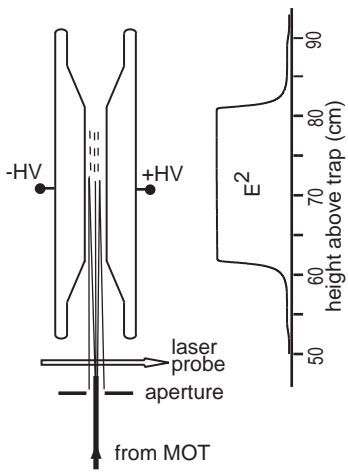}
\caption{\label{apparatus} Schematic of  the fountain apparatus.   Atoms launched from the trap are: bunched to reduce their velocity spread (the buncher is not shown), collimated by an aperture, and fluoresced by a probe laser beam either while rising or falling.  A profile of the electric field squared ($E^2$) is shown to the right. The tapered field plate entrance reduces defocusing effects \cite{Maddi99PRA60p3882}, while the larger gaps at the ends are to provide a weaker field for other experiments.
}
\end{figure}

\begin{figure}
\epsfig{file=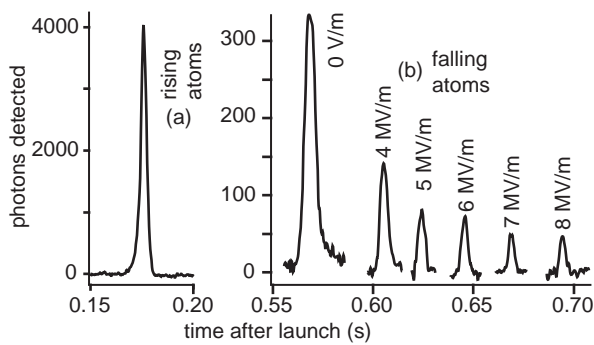}
\caption{\label{arrival} Fluorescence signals for atoms (a) rising into the plates and (b) falling from the plates as a function of time after launch. Plot (b) is a composite of observed return signals for zero electric field and five of the nonzero electric fields used in the experiment.  The vertical axis is the number of photons detected in a 0.5 ms interval.  
}
\end{figure}

In our measurement, cesium atoms are launched vertically from a magneto optic trap (MOT) such that the atoms reach their zenith between a set of parallel electric field plates (Fig.~\ref{apparatus}). When the electric field is turned on, the neutral cesium atoms accelerate into the plates  with a resulting  higher trajectory and a correspondingly longer  flight time (Fig.~\ref{arrival}).  The kinetic energy boost afforded the atoms by the electric field is then determined from the increase in round trip time.

%
%

%
%

The atoms are detected by laser-induced fluorescence just before they enter the electric field plates and again when they fall back out.  The intensity of the fluorescence is recorded against time as a measure of the cesium packet density profile.  The height of the laser probe was determined to within 0.1~mm with respect to the field plates using  reference features fixed to the electric field plate assembly and surveyed as part of the measurement of the electric field plate gap as discussed below.  Because the laser fluorescence is destructive, the laser beam was shuttered to allow detection of the atoms either when they are rising or when they are falling.

The electric field plates were machined from aluminum with the high-field surfaces tungsten-coated for sputter resistance. The gap between the electric field plates (3.979~mm average) was measured to a precision of $\pm 1~\mu$m (dominated by shot noise) along its length and width by profiling the  interior surfaces with a 5-axis coordinate measuring  machine (Fanamation model 606040). A calibration correction of  $-10.9\pm0.3~\mu$m was determined by profiling a set of precision gaps constructed of sandwiched gauge blocks.   Variations of 4~$\mu$m across the width of the plates were averaged with a weighting factor favoring the midline of the plates, where the atoms spend the greatest portion of the round trip. The resulting electric fields were calculated  for the full length of the plates using a two-dimensional finite-element code. The voltages applied to the electric field plates were continuously measured by two NIST-traceable voltage dividers (Ross model VD30-8.3-BD-LD-A) and a load matched multimeter (HP3457A).

A single measurement of the polarizability consisted of recording the intensity versus time of the fluorescence signals from two rising and two falling cesium atom packets in zero electric field, followed by  five falling packets with the electric field on.  The polarizability required to generate the delay in the falling cesium packet with the electric field  on was determined by integrating the equations of motion over the path of the packet.  The local value for gravity ($979.92\pm0.03$~cm/s$^2$) was interpolated  from the local gravity measurements in Ref. \cite{Ponce01USGS}. 

\begin{figure}
\epsfig{file=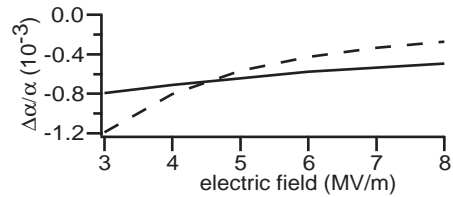}
\caption{\label{corrections} Corrections applied to the  measured polarizability to compensate for the longitudinal velocity width of the packet (solid line) and non-uniform losses (broken line). The velocity width correction has a \per{13} uncertainty and the loss correction has a \per{100} uncertainty.}
\end{figure}

The resulting polarizabilities  were corrected for the longitudinal velocity spread of the cesium packet.  To minimize this correction, we used a version of the pulsed electric-field technique presented in Ref.~\cite{Maddi99PRA60p3882} to reduce the longitudinal velocity spread of the cesium packet shortly after launch from 3.2~cm/s RMS to 0.8~cm/s.    The reduction in the velocity spread also serves to decrease the time width of the fluorescence signal when detecting the cesium packet and consequently increase our time resolution.  The corrections for the velocity width are field dependent and are shown in Fig.~\ref{corrections}. The average correction is \per{-0.07}.

%
%

Non-uniform losses across the longitudinal width of the packet distort the shape of the fluorescence signal.  Numerical evolution of the longitudinal and transverse phase-space of the packet through the apparatus, taking into account sources of defocusing and clipping, generated the non-uniform loss corrections to the polarizability shown in Fig.~\ref{corrections}. The average correction was \per{-0.05}.

\begin{figure}
\epsfig{file=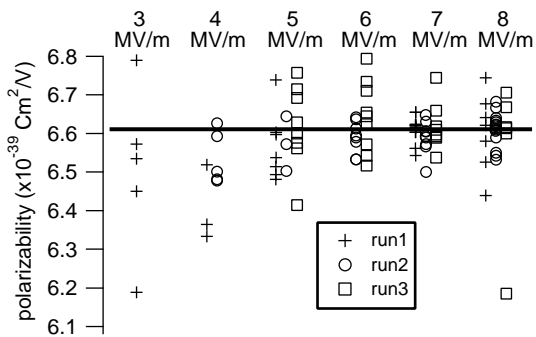}
\caption{\label{allalpha} All of the polarizability data points for each of the six electric fields. The data were collected in three runs as indicated in the legend.  The runs are shown  displaced from each other along the horizontal axis for clarity.  For lower voltages, the change in transit time is smaller, resulting in lesser sensitivity and a consequently larger scatter in $\alpha_0$.
}
\end{figure}

\begin{figure}
\epsfig{file=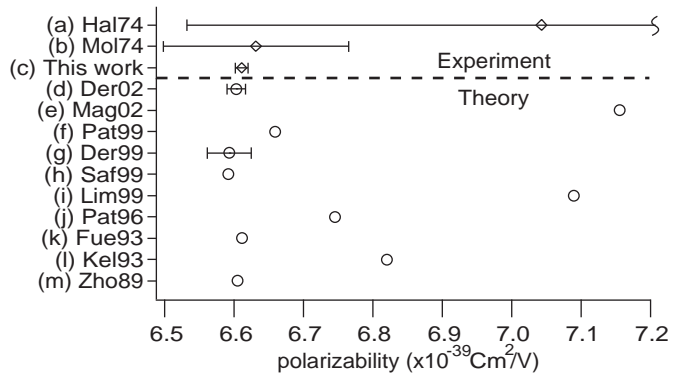}
\caption{\label{alpharesults} The most recent measurements and calculations of the cesium static dipole polarizability.  The top three points represent measurements.
(a) is Ref. \cite{Hall74PRA10p1141}, 
(b) is Ref. \cite{Molof74PRA10p1131}, 
(c) is this work,
(d) is Ref. \cite{Derevianko02PRA65p053403}, 
(e) is Ref. \cite{Magnier02JQSRT75p121}, 
(f) is Ref. \cite{Patil99CPL301p64}, 
(g) is Ref. \cite{Derevianko99PRL82p3589}, 
(h) is Ref. \cite{Safronova99PRA60p4476}, 
(i) is Ref. \cite{Lim99PRA60p2822}, 
(j) is Ref. \cite{Patil96JCP106p2298},
(k) is Ref. \cite{Fuentealba93JPB26p2245}, 
(l) is Ref. \cite{Kello93PRA47p1715}, and
(m) is Ref. \cite{Zhou89PRA40p5048}.
Uncertainties are shown when included in the publication.}
\end{figure}

The data, totaling 105 measurements of $\alpha_0$, were taken in three runs and over electric fields of 3 MV/m  to 8 MV/m (Fig.~\ref{allalpha}). The final result is  $\alpha_0(\mbox{Cs}) = 6.611\pm 0.009\times 10^{-39}~\mbox{Cm}^2/\mbox{V}= 59.42\pm 0.08\times 10^{-24}~\mbox{cm}^3=401.0\pm0.6~a_0^3$. The error budget is summarized in Table~\ref{stats}.  Our value for the polarizability is plotted in Fig.~\ref{alpharesults} along with those from previous measurements and from recent calculations. 

%
%
%
\begin{table}
\caption{\label{stats} Sources of error.}
\begin{ruledtabular}
\begin{tabular}{lcd}
Source & uncertainty & \multicolumn{1}{c}{\mbox{\% of $\alpha_0$}} \\
\hline
Calibration of 4 mm gap measurement& 0.3 $\mu$m & 0.015 \\
Thermal expansion of the 4 mm gap & 0.8 $\mu$m & 0.04 \\
Fits to plate shapes &1 $\mu$m & 0.05 \\
Velocity width and atom losses & ------ & 0.05 \\
Laser probe height uncertainty & 0.1 mm & 0.05 \\
Deviation from vertical & 5 mrad & 0.0026 \\
Gravitational acceleration & 0.03 cm/s$^2$ & 0.026 \\
Path integration errors & 0.01 ms & 0.01 \\
Finite width of the electric field & 0.03 ms & 0.03 \\
Defocusing on exit of field plates & ----- & 0.025 \\
Stray magnetic field gradients& ----- & 0.013 \\
Voltage divider ratio &\per{0.01} & 0.028 \\
Divider load matching & 0.02 M$\Omega$ & 0.040 \\
Voltage measurements & 0.1 mV& 0.030 \\
\hline
Total systematic uncertainty & & 0.12 \\
Statistical uncertainty & & 0.065 \\
Total uncertainty & & 0.14\\
\end{tabular}
\end{ruledtabular}
\end{table}
%
%
%

\begin{figure}
\epsfig{file=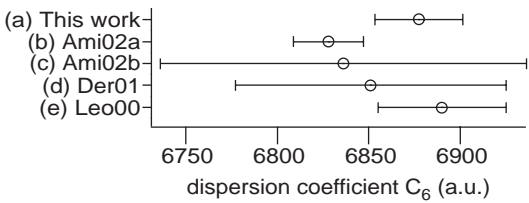}
\caption{\label{c6comparison} Comparison of recent values of $C_6$ for cesium. 
(a) is this work with Ref. \cite{Derevianko02PRA65p053403},
(b) is Ref. \cite{Amiot02PRA66p052506},
(c) is Ref. \cite{Amiot02JCP117p5155},
(d) is Ref. \cite{Derevianko01PRA63p052704}, and
(e) is Ref. \cite{Leo00PRL85p2721}.
}
\end{figure}

The recent calculation of $\alpha_0$ and the cesium $6^2P_{1/2}$ and $6^2P_{3/2}$ lifetimes from the dispersion coefficient $C_6$  by Derevianko and Porsev \cite{Derevianko02PRA65p053403} demonstrates the intimate connection between these quantities. Reversing their procedure, we extract from our value of $\alpha_0$ the contribution of the $6^2P_{1/2}$ and $6^2P_{3/2}$ states to the ground state polarizability. These two states account for \per{96} of $\alpha_0$.  With the ratio of dipole matrix elements measured in Ref.~\cite{Rafac98PRA58p1087}, we obtain the absolute value of the  $6^2S_{1/2}$ to $6^2P_{1/2}$ and $6^2S_{1/2}$ to $6^2P_{3/2}$ reduced dipole matrix elements $|D_{1/2}| = (3.824\pm0.003\pm0.001)\times10^{-29}~\mbox{C\,m} = 4.510\pm0.004\pm0.001~\mbox{a.u.}$ and $|D_{3/2}| = (5.381\pm0.004\pm0.001)\times10^{-29}~\mbox{C\,m} = 6.347\pm0.005\pm0.002~\mbox{a.u.}$, respectively, where the reduced dipole matrix elements are defined according to the Wigner-Eckart theorem formulated with 3-J symbols \cite{AMOHandbook96}. The definitions for  atomic units (a.u.) are given in Ref.~\cite{AMOHandbook96}. The values listed here have their uncertainties separated into two portions that are displayed in the form $\pm\sigma_1\pm\sigma_2$ where $\sigma_1$ is the contribution from the uncertainty in our value of $\alpha_0$ and $\sigma_2$ is the contribution from those values provided by Ref.~\cite{Derevianko02PRA65p053403}. The total uncertainty is the sum of these two values in quadrature. From the reduced dipole matrix elements, we obtain lifetimes \cite{Derevianko02PRA65p053403,Sakurai67AdvQuantMech} for the $6^2P_{1/2}$ and $6^2P_{3/2}$  states of $34.72 \pm 0.06 \pm 0.02~\mbox{ns}$ and $30.32 \pm 0.05 \pm 0.02~\mbox{ns}$, respectively.  Using the relation between  $D_{1/2}$ and $C_6$ given by Derevianko and Porsev, we obtain $C_6 = (6.584\pm0.020\pm0.012)\times10^{-76}~\mbox{J\,m}^6 = 6877\pm21\pm12~\mbox{a.u.}$.  Our values are compared with other determinations of the $6^2P_{1/2}$ and $6^2P_{3/2}$ lifetimes in Table~\ref{lifetimes} and of $C_6$ in Fig.~\ref{c6comparison}.   Our values for the lifetimes agree with the calculation of Ref.~\cite{Derevianko02PRA65p053403} and the measurements of  Refs.~\cite{Young94PRA50p2174, Rafac94PRA50pR1976} but differ from the values given in Refs.~\cite{Amiot02PRA66p052506,Rafac99PRA60p3648}.  

%
\begin{table}
\caption{\label{lifetimes} Cesium $6P_{1/2}$ and $6P_{3/2}$ lifetimes in ns.}
\begin{ruledtabular}
\begin{tabular}{ll@{$\pm$}ll@{$\pm$}ll}
method & \multicolumn{2}{c}{$6^2P_{1/2}$} & 
	\multicolumn{2}{c}{$6^2P_{3/2}$} & reference\\
        \hline
from $\alpha_0$ & 34.72 & 0.06 & 30.32 & 0.05 & 
	this work with Ref. \cite{Derevianko02PRA65p053403}\\
from PAS$^*$ & 34.88 & 0.02 & 30.462 & 0.003 &
	Amiot et. al. \cite{Amiot02PRA66p052506}\\
from $C_6$ & 34.80 & 0.07 & 30.39 & 0.06 &
	Derevianko \& Porsev \cite{Derevianko02PRA65p053403}\\
meas. & 35.07 & 0.10  & 30.57 & 0.07 & 
	Rafac et. al (1999) \cite{Rafac99PRA60p3648}\\
meas.  & 34.934 & 0.094 & 30.499 & 0.070  & 
	Rafac et. al (1994)\cite{Rafac94PRA50pR1976}\\
meas. & 34.75 & 0.07 & 30.41 & 0.10  & 
	Young et. al (1994) \cite{Young94PRA50p2174}\\
\end{tabular}
\end{ruledtabular}
* PAS: photoassociative spectroscopy.
\end{table}
%
%
%


In conclusion, from the change in the time-of-flight of a fountain of neutral Cs atoms passing through a uniform electric field, we have determined the static scalar dipole polarizability of the cesium $6^2S_{1/2}$ ground state to an uncertainty of \per{0.14}.  This is sufficient to test high precision calculations that include core electron contributions. From our polarizability result, we have derived the lifetimes of the cesium $6^2P_{1/2}$ and $6^2P_{3/2}$ states and the cesium-cesium dispersion coefficient $C_6$.
 

We thank Timothy Page, Daniel Schwan,  Xinghua Lu, and  Xingcan Dai for their assistance in the construction and assembly  of the experiment, Robert Connors for his  help in profiling the electric field plates, and Timothy P. Dinneen for his development of the prototype of this fountain apparatus.   This work was supported by the Office of Biological and Physical Research, Physical Sciences Research Division, of the National Aeronautics and Space Administration and in its early stages by the  Office of Science, Office of Basic Energy Sciences,  of the U.S. Department of Energy, under Contract  No. DE-AC03-76SF00098. One of us (JA) acknowledges support from the NSF and from NASA.

%
%
\bibliography{polarizability,MasterBibliography}

\end{document}